\documentclass[10pt]{article}
\usepackage{preamble}
\title{Droplet formation simulation using mixed finite elements\footnote[1]{This article may be downloaded for personal use only. Any other use requires prior permission of the author and AIP Publishing. This article appeared in~\cite{NathawaniKnepleyDropletGravity} and may be found at \url{https://doi.org/10.1063/5.0089752}}} 
\author[1]{\textbf{Darsh K. Nathawani}\footnote[2]{Email address for correspondence: darshkir@buffalo.edu}}
\affil[1]{Computational and Data-Enabled Science and Engineering, University at Buffalo, Buffalo, NY 14260, USA}
\author[2]{\textbf{Matthew G. Knepley}}
\affil[2]{Department of Computer Science and Engineering, University at Buffalo, Buffalo, NY 14260, USA}

\begin{document}
\makeatletter
\hfil\parbox[t]{0.7\textwidth}{\centering\LARGE\bfseries\@title}\par
\vspace{1cm}
\hfil\parbox[t]{0.7\textwidth}{\centering\bfseries\@author\\[3ex]\@date}\par
\makeatother
\begin{abstract}
Droplet formation happens in finite time due to the surface tension force. The linear stability analysis is useful to estimate droplet size but fails to approximate droplet shape. This is due to a highly non-linear flow description near the point where the first pinch-off happens. A one-dimensional axisymmetric mathematical model was first developed by Eggers and Dupont\cite{EggersDupont1994} using asymptotic analysis. This asymptotic approach to the Navier-Stokes equations leads to a universal scaling explaining the self-similar nature of the solution. Numerical models for the one-dimensional model were developed using the finite difference\cite{EggersDupont1994} and finite element method\cite{AmbravaneswaranWilkesBasaran2002}. The focus of this study is to provide a robust computational model for one-dimensional axisymmetric droplet formation using the Portable, Extensible Toolkit for Scientific Computation (PETSc). The code is verified using the Method of Manufactured Solutions (MMS) and validated using previous experimental studies done by Zhang and Basaran\cite{ZhangBasaran1995}. The present model is used for simulating pendant drops of water, glycerol, and paraffin wax, with an aspiration of extending the application to simulate more complex pinch-off phenomena.  
\end{abstract} 
\section{{\label{sec:1}Introduction}}
Singularity in free surface flows is a crucial problem for time-accurate simulations. A mathematical treatment, built on the self-similarity of the pinch-off region, has made numerical simulations tractable. Our study is concerned with the formation of pendant droplets. Rayleigh was the first to demonstrate that droplet formation occurs in finite time due to the force of surface tension acting against inertia~\cite{Rayleigh1878}. The pinch-off dynamics of a pendant drop has received the most attention from both mathematicians and scientists. Considering the pendant drop as a fluid column, the surface tension force and the gravitational force are initially balanced. The drop becomes heavier as more fluid is added from the top. Eventually gravity extends the drop, increasing the surface energy. In order to minimize this energy, the radius of the fluid column shrinks, forming a neck region, what we see as the action of surface tension. At some finite time, the radius becomes zero at some location and the drop separates from the fluid column. This location in the fluid column is called the \textit{pinch-off point} or singularity.

Fluid motion in the immediate vicinity of the singularity is driven by very high velocity gradients generated by the surface tension, inertial, and viscous forces. In fact, the solution in this region is self-similar, meaning that it does not depend on the initial or boundary conditions, but has a universal character. After the first pinch-off, a long neck recoils back with high velocity. This induces surface perturbations in the column and can lead to further breakup into smaller \textit{satellite} droplets, a phenomenon also observed in liquid bridges and decaying jets. A now-classic treatment of the governing dynamics of singularities as well as analysis of the self-similarity for these cases is given by Eggers~\cite{Eggers1997, Eggers1995}. Linear stability analysis can accurately approximate droplet size, but fails to approximate the shape of a droplet~\cite{ChaudharyRedekopp1980a}. Moreover, even the higher order analysis is not able to explain the shape of the drop near singularity~\cite{ChaudharyRedekopp1980a, ChaudharyMaxworthy1980b}. One-dimensional analysis has proved useful for circular liquid jets~\cite{Bogy1979,Lee1974} and pendant drops~\cite{Cram1984}. On the experimental side, studies have examined the pinch-off dynamics of pendant drops~\cite{Peregrine1990, ShiBrennerNagel1994}, as well as characterized droplet dynamics in terms of non-dimensional parameters~\cite{ZhangBasaran1995}.

A full one-dimensional mathematical model was constructed by Eggers and Dupont~\cite{EggersDupont1994} using the asymptotic expansion of the Navier-Stokes equations in cylindrical coordinates. They used a finite difference scheme to discretize the equations, and simulated both a pendant drop and a decaying jet. They verified that one-dimensional treatment can accurately simulate the pinch-off dynamics for a fluid in a quiescent background. However, their computational approach could only simulate up to the first pinch-off. Other computational models using one and two-dimensional analysis were explored by Ambravaneswaran et al.~using finite elements to discretize the problem~\cite{AmbravaneswaranWilkesBasaran2002}. They investigated the effect of volume flow rate on droplet and neck shape. Moreover, they were able to simulate satellite drops. However, they were not able to validate the satellite drop simulations, as they had the primary droplet. Their two-dimensional simulations support the conclusion that for axisymmetric droplet formation, one-dimensional computational models are much faster and reliably accurate. Ambravaneswaran et al.~also propose a hybrid 1D-2D computational model, and matching between 1D and 2D domains is also explored~\cite{KistlerScriven1984}. Other numerical approaches like the Volume-of-Fluid (VOF) method were tested, but either failed to accurately capture features of the flow, such as micro-threads, or were limited to only a certain range of fluid parameters, such as viscosity~\cite{RichardsBerisLenhoff1995,Zhang1999}. Additionally, the VOF method requires at least a 2D domain, which increases the computational cost.

Formation of droplets can be seen in many scientific and industrial applications, a few examples being ink-jet printing~\cite{Derby2010}, spray cooling~\cite{Kim2007}, and droplet entrainment in annular flow~\cite{BernaEscrivaMunozHerranz2015}. Moreover, droplet formation has a pivotal role in fuel entrainment and burning in hybrid rockets~\cite{KarabeyogluCantwell2002}, which is the motivation to pursue this study. We present a computational model that is both verified with MMS and validated with previous experimental work. We extend the results with paraffin wax simulations to build a base for the future work on droplets in a shear force environment.
\section{\label{sec:2}Mathematical and Computational Model}
\begin{figure}[t]
    \centering
    \includegraphics[width=0.6\textwidth]{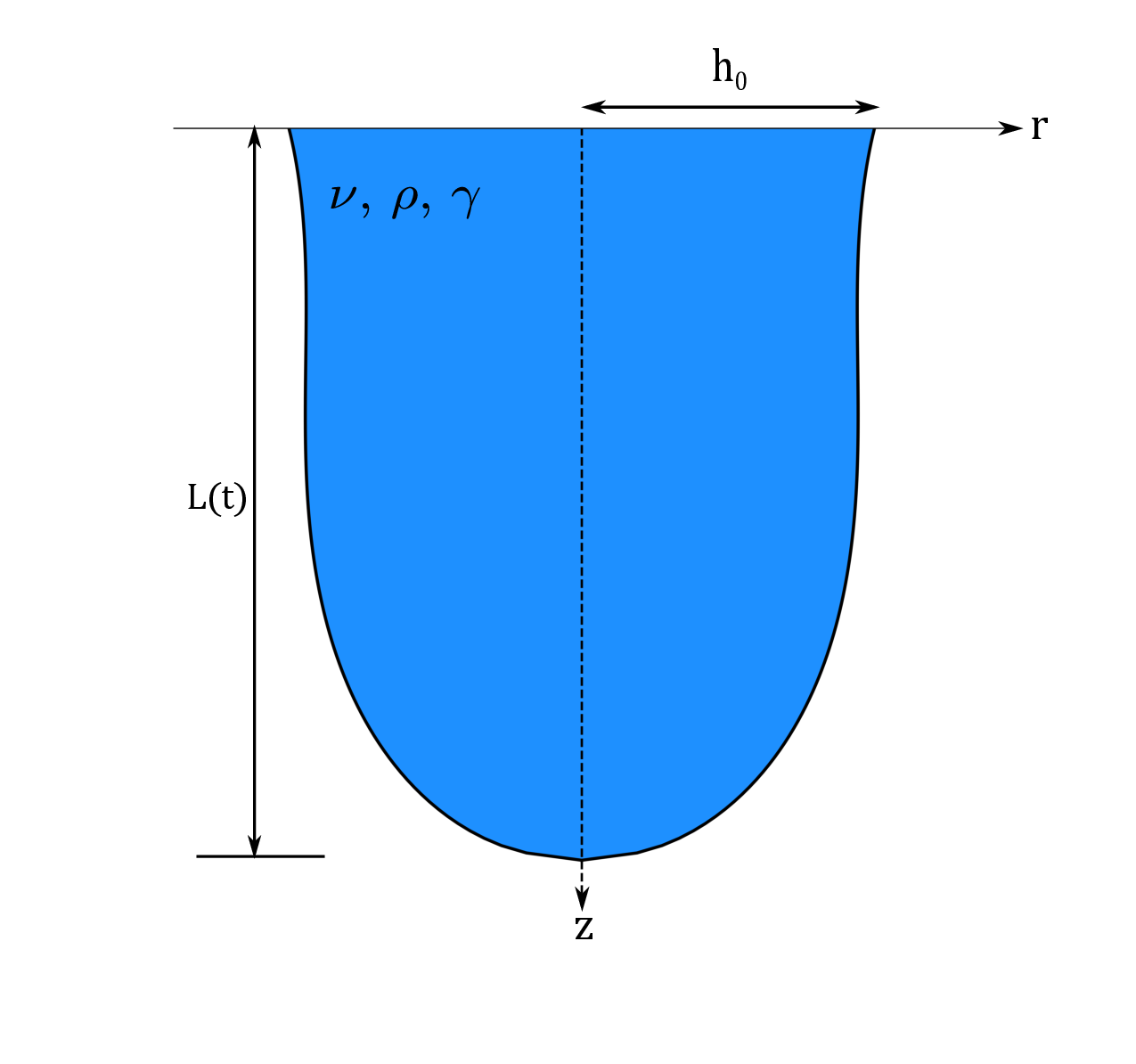}
    \caption{Schematic of a pendant droplet.}
    \label{fig:schematic}
\end{figure}

In this section, we consider the Navier-Stokes momentum equation in cylindrical coordinates for an axisymmetric fluid column, as treated by Eggers and Dupont~\cite{EggersDupont1994}. The fluid is considered incompressible with density $\rho$ and kinematic viscosity $\nu$. Assuming no swirling motion, we consider the flow only in radial and axial directions. A schematic of a pendant drop in cylindrical coordinates is shown in Fig.~\ref{fig:schematic}. The surface of the droplet, which is defined as variable $h(z,t)$, is moving with the velocity. Therefore, the model equation for $h$ is given by
\begin{align*}
    \frac{\partial h}{\partial t} + u_z \frac{\partial h}{\partial z} = u_r \vert_{r=h}
\end{align*}

Approaching the pinch-off point, the radial contraction is much faster than the axial expansion. Therefore, considering the radius $r$ as an asymptotic parameter, the axial velocity ($u_z$) and pressure ($p$) are expanded asymptotically in even order terms to satisfy symmetry. Then using the continuity equation, the radial velocity is derived.
\begin{align*}
  u_{z} &= u_{0} + u_{2} r^{2} + \ldots, \\[1ex]
  u_{r} &= - \frac{\partial u_0}{\partial z} \frac{r}{2} - \frac{\partial u_2}{\partial z} \frac{r^{3}}{4} - \ldots, \\[1ex]
  p   &= p_{0} + p_{2} r^{2} + \ldots
\end{align*}

To introduce the surface tension force ($\gamma$) in the governing equations, the following force balance is considered.
\begin{align*}
    \mathbf{\hat{n}} \sigma \mathbf{\hat{n}} &= -\gamma \left(\nabla \cdot \mathbf{\hat{n}} \right) \\[1ex]
    \mathbf{\hat{n}} \sigma \mathbf{\hat{t}} &= 0
\end{align*}
Here, $\sigma$ is a stress tensor, $\mathbf{\hat{n}}$ is a unit outward normal, $\mathbf{\hat{t}}$ is a unit tangent, and $\nabla \cdot \mathbf{\hat{n}}$ is a mean curvature. The above force balance explains that the normal stress is balanced by the surface tension and the tangential stress is zero.

Using the force balance and the leading order terms in $r$ from the expansion, we simplify the momentum equation. The advecting surface equation is already in leading order. Dropping the subscripts, the governing equations for a one-dimensional axisymmetric fluid column are given by
\begin{align}
    \frac{\partial u}{\partial t} +  u \frac{\partial u}{\partial z} + \frac{\gamma}{\rho} \frac{\partial (\nabla \cdot \mathbf{\hat{n}})}{\partial z} - \frac{3 \nu}{h^2}\frac{\partial}{\partial z} \left( h^2 \frac{\partial u}{\partial z}\right) - g &= 0 \label{eq:1} \\[2ex]
    \frac{\partial h}{\partial t} + u \frac{\partial h}{\partial z} + \frac{h}{2}\frac{\partial u}{\partial z} &= 0
    \label{eq:2}
\end{align}
where, the mean curvature term $\nabla \cdot \mathbf{\hat{n}}$ is given by
\begin{align}
  \nabla \cdot \mathbf{\hat{n}} = \left[ \frac{1}{h \left (1 + \frac{\partial h}{\partial z}^2\right )^{1/2}} - \frac{\frac{\partial^2 h}{\partial z^2}}{\left (1 + \frac{\partial h}{\partial z}^2\right )^{3/2}} \right]
\end{align}
Here, the curvature term is not approximated to the leading order because it has been shown that using the full curvature term better captures the singularities~\cite{Eggers1995}.

Equations (\ref{eq:1}) and (\ref{eq:2}) govern a pendant drop under gravitational forcing. These are solved for $u$ and $h$ using a finite element discretization. However, the highest order derivative is of the third order, which is problematic for our $C^0$ continuous element scheme. The approximation for this term will be discontinuous across element interfaces. We could handle this using a discontinuous Galerkin (DG) scheme, but instead, we choose a mixed-element formulation, inspired by Ambravaneswaran et~al.~\cite{AmbravaneswaranWilkesBasaran2002}, in which we explicitly discretize the axial derivative of the radius $h$ (or slope), $s = \partial h / \partial z$, so that
\begin{align}
  \nabla \cdot \mathbf{\hat{n}} = \left[ \frac{1}{h \left (1 + s^2\right )^{1/2}} - \frac{\frac{\partial s}{\partial z}}{\left (1 + s^2\right )^{3/2}} \right] \label{eq:4}
\end{align}
The mixed finite element formulation is given by 
\begin{align}
\int_{\Omega} q \left[ \frac{\partial u}{\partial t} +  u \frac{\partial u}{\partial z} + \frac{\gamma}{\rho} \frac{\partial (\nabla \cdot \mathbf{\hat{n}})}{\partial z} - \frac{3 \nu}{h^2}\frac{\partial}{\partial z} \left( h^2 \frac{\partial u}{\partial z}\right) - g \right] d \Omega  &= 0 && \label{eq:5} \\[2ex]
\int_{\Omega} v \left [ \frac{\partial h}{\partial t} + u \frac{\partial h}{\partial z} + \frac{1}{2} h \frac{\partial u}{\partial z} \right ]d \Omega &= 0 && \label{eq:6} \\[2ex]
\int_{\Omega} w \left [ s - \frac{\partial h}{\partial z} \right ] d \Omega &= 0 && \label{eq:7}
\end{align}
Here, $q$, $v$, and $w$ are test functions, and the mean curvature is defined by Eq.~(\ref{eq:4}). The third and fourth terms in Eq.~(\ref{eq:5}) are simplified by performing integration by parts, which makes the weak form with the highest order derivative to first order. The equations after integration by parts take the following form.
\begin{align}
\int_{\Omega} q \left[ \frac{\partial u}{\partial t} +  u \frac{\partial u}{\partial z} -\frac{6\nu}{h}\frac{\partial h}{\partial z}\frac{\partial u}{\partial z} + \frac{\gamma}{\rho} \left\{-\frac{s \frac{\partial s}{\partial z}}{h \left (1 + s^2\right )^{3/2}} - \frac{s}{h^2 \left (1 + s^2\right )^{1/2}}\right\} - g \right] d\Omega \nonumber \\[1ex]
+ \int_{\Omega} \nabla q \left[3 \nu \frac{\partial u}{\partial z} + \frac{\gamma}{\rho} \frac{\frac{\partial s}{\partial z}}{\left (1 + s^2\right )^{3/2}}  \right] d \Omega - \int_{\Gamma} \nabla q \left[3 \nu \frac{\partial u}{\partial z} + \frac{\gamma}{\rho} \frac{\frac{\partial s}{\partial z}}{\left (1 + s^2\right )^{3/2}}  \right] d \Omega  &= 0 && \label{eq:8} \\[2ex] 
\int_{\Omega} v \left [ \frac{\partial h}{\partial t} + u \frac{\partial h}{\partial z} + \frac{1}{2} h \frac{\partial u}{\partial z} \right ]d \Omega &= 0 && \label{eq:9} \\[2ex]
\int_{\Omega} w \left [ s - \frac{\partial h}{\partial z} \right ] d \Omega &= 0 \label{eq:10} && 
\end{align}

Initially, the velocity is zero and the curvature profile is a hemisphere as it minimizes surface energy. The inlet radius $h_0$ is fixed depending on the nozzle radius and the inflow velocity $u_0$ is constant. The radius at the tip of the droplet, at length $L(t)$, is zero. The set of Eq.~(\ref{eq:8})-(\ref{eq:10}) are then solved using a continuous Galerkin formulation subject to the following constraints.\\[2ex]
\noindent\textbf{Initial conditions:}
\begin{align*}
    h &= \sqrt{h_0^2 - z^2} \\[1ex]
    s &= -\frac{z}{\sqrt{h_0^2 - z^2}}\  \text{for}\ (0\leq z < L_0),\qquad s|_{L_0} = - C \\[1ex]
    u &= 0
\end{align*} 
where C is a large negative number. In our implementation, we use $-10$. However, the code was tested with larger values and the results were unchanged.\\[2ex]
\noindent\textbf{Boundary conditions:}
\begin{center}
\setlength{\tabcolsep}{12pt}
\begin{tabular}{l | l l}
 $z = 0$   & $h = h_0$ & $u = u_0$ \\[1.5ex]
 $z = L(t)$  & $h = 0$   & $u = \frac{dL}{dt}$
\end{tabular}
\end{center}

The length of the drop L(t) can be calculated as a part of the solution as explained by Ambravaneswaran et al.~\cite{AmbravaneswaranWilkesBasaran2002} by calculating the volume of the drop, which can then be used to calculate the velocity at the tip. However, this results in a dense row in the Jacobian, so we instead produce $L(t)$ by self-consistent iteration.

Initially, we are given $u(t)$ and $h(t)$, including the velocity at the end of the droplet.
We use that velocity to predict $L(t + dt)$ that is $L(t) + u_{tip}*dt$, giving us our boundary condition, and we extend our mesh to this length. 
We then solve our existing system (Eq.~(\ref{eq:8})-(\ref{eq:10})) for $u(t + dt)$ and $h(t + dt)$.
This allows us to calculate the droplet volume by integrating $h(t + dt)$ along the length.
This must match the volume from the last time-step augmented by the amount of liquid flowing, which is $4 \pi h^2_0 u_0 dt$.
The difference between the calculated volume and the theoretical volume is used for self-consistency in adjusting the length $L(t + dt)$.
We use bisection to arrive at a consistent length $L$ for this new time step.
This adaptation loop is done when the conservation of volume is satisfied to a given tolerance (we use 0.1\%).

\begin{figure}[t]
    \centering
   \includegraphics[width=0.4\textwidth]{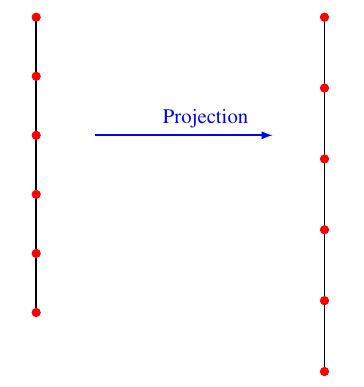}
    \caption{Scaling of the mesh according to $L(t)$.}
    \label{fig:projection}
\end{figure}

The one-dimensional mesh, representing the domain $0 \leq z \leq L(t)$, moves as the length $L(t)$ changes. We first update the position of the last vertex and then move the remaining vertices to even out the cell lengths. The interpolation of the discrete field representation between these two meshes can be achieved using the Galerkin projection~\cite{FarrellMaddison2011}. Galerkin projection is optimal for the $L_2$ norm, which measures energy. Alternatively, we could replace this with a volume constraint during interpolation. As shown in Fig.~\ref{fig:projection}, the re-meshing is done using the calculated $L(t)$ between each time step. The Galerkin projection is then used to interpolate the solution to the new mesh. For instance, if the solution on the old mesh is $u^{old}$ and on the new mesh is $u^{new}$, then the interpolation is done as follows:
\begin{align*}
u^{new} &= \eta^{new}_i \left (u^{old}\right ) \\
u^{new} &= \eta^{new}_i \left (\sum_k u^{old}_k \phi^{old}_k \right ) \\
u^{new} &= \sum_k u^{old}_k \eta^{new}_i \left (\phi^{old}_k \right ) \\
\intertext{where,}
\eta^{new}_i \phi^{old}_k &= \sum_q w_q \phi^{old}_k(x_q)    
\end{align*}
Here, $\phi$ represents the basis, $\eta$ represents the dual basis, $x_q$ are the quadrature points and $w_q$ are the weights on the quadrature points. 
\begin{algorithm}[t] 
\caption{Self consistency using TSAdapt} 
\label{algo:1} 
\begin{itemize}
\small
\item \textbf{PreStep:}
\begin{itemize}
\item Calculate $ V_{t+1} = V_t + V_{dt} $ ($V_0$ is known)
\item Calculate $ l_{t+1} = l_t + (u_{tip}*dt) $
\item Scale the mesh and Project the solution 
\item Label the cells to refine
\end{itemize}
\item \textbf{TSSolve:} {Solve the equations}
\item \textbf{TSAdapt:}
\begin{itemize}
\item Calculate $V_{new}$
\item \textbf{If}($V_{new} = V_{t+1}$) $\Rightarrow$ Accept the solution
\item \textbf{Else}
\begin{itemize}
\item[-] Reject the solution 
\item[-] Adapt the length using $V_{new}$ and $V_{t+1}$ 
\item[-] Scale the mesh and Project the solution 
\item[-] GoTo: \textbf{TSSolve}
\end{itemize}
\end{itemize}
\item \textbf{PostStep:} { Refine the labeled cells}
\end{itemize}
\end{algorithm}

Calculating the length, scaling the mesh, projecting the solution and the length adaptation (when volume lost is more than a specified threshold) can be merged into a self-consistent loop as shown in Algorithm~\ref{algo:1}. The neck requires sufficiently refined mesh to capture the singularity. Hence, we start with a coarse mesh and refine it as we approach the singularity. The elements are labeled for refinement based on the radius and velocity gradients. Before the next time step, the labeled elements are then refined if necessary. This adaptive mesh refinement is also included in the algorithm~\cite{LangeMitchellKnepleyGorman2015,WallworkKnepleyBarralPiggott2022}. We use the Portable, Extensible Toolkit for Scientific Computation (PETSc)~\cite{petsc-user-ref,petsc-web-page} to set up and solve the system using time-stepper (TS) object~\cite{tspaper}. The self-consistent algorithm is set up using \textit{TSAdapt} functionality of TS. We use a direct solver using LU factorization.
\section{\label{sec:3}Results and discussion}
In this section, we discuss the verification and validation of the numerical model presented in the previous section. Then, we explore the pinch-off dynamics in paraffin wax.
\subsection{Verification and Validation}
\begin{table}[b]
\caption{\label{table:1}Physical properties of the materials used for the simulations.}
\vspace{2mm}
\renewcommand{\arraystretch}{1.5} 
\centering
    \begin{tabular}{lccc}
\hline
\hline
Material & $\rho$ (kg/m$^3$) & $\nu$ (m$^2$/s) & $\gamma$ \footnote{Surface tension value at liquid-air interface.}  (N/m) \\
\hline 
\hline
Water    & $998$  &  $1.0 \times 10^{-6}$ & $0.0728$      \\
$85\%$ Glycerol & $1223$ & $9.2 \times 10^{-5}$ & $0.066$   \\
Paraffin wax \footnote{Wax properties are at 100$^{\circ}$ C.}   & $760$ & $4.2 \times 10^{-6}$ & $0.028$ \\
\hline
\end{tabular}

\end{table}

\begin{figure}[t]
    \centering
    \includegraphics[width=0.5\textwidth]{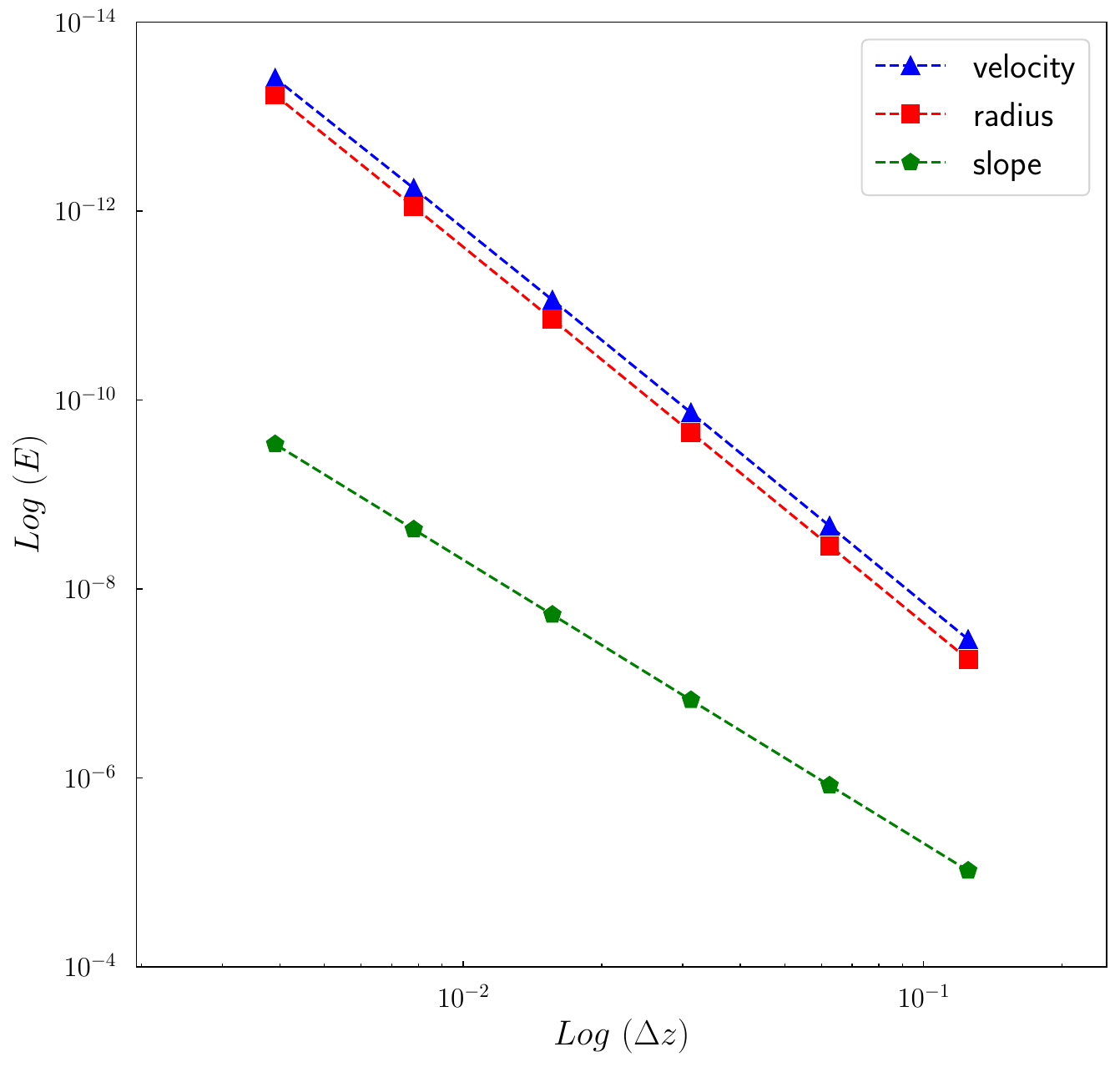}
    \caption{Discretization errors in velocity ($u$), radius ($h$) and slope ($s$) for different mesh sizes using manufactured solution.}
    \label{fig:error_plot}
\end{figure}

Before proceeding with the computational model, it is vital to perform a verification test. Verification is a mathematical exercise that can be used to examine the error evaluation done by the implemented code. One elegant method for code verification is the Method of Manufactured Solution (MMS)~\cite{Roache2002}. This is a very straightforward method, where we simply pick a non-trivial solution and add the source term into the equation generated by applying the operator on the solution. This way we know the exact solution of the modified equation (original equation with the source term). Then the error evaluation for this modified equation must be zero. The computational model we use is verified using the MMS. The MMS helps to eliminate coding errors and is also useful to test the discretization for problems with unknown exact solutions. Figure~\ref{fig:error_plot} shows log-log plot of $L_2$ norm of the error, $|| u_{fe} - u_{mms} ||_{L_2}$, where $u_{fe}$ is finite element solution and $u_{mms}$ is the manufactured solution. The velocity ($u$) and radius ($h$) are discretized using third-order polynomials, whereas slope ($s$) is discretized using second-order polynomials. The error reduces by order four for $u$ and $h$, and order three for $s$, verifying the correct implementation of the numerical model. The error evaluation is done on a moving mesh with the scaling factor of $1.0001$. The MMS solution is evaluated every time step on a scaled mesh and compared to the solution.
\begin{figure}[!b]
    \centering
    \includegraphics[width=0.7\textwidth]{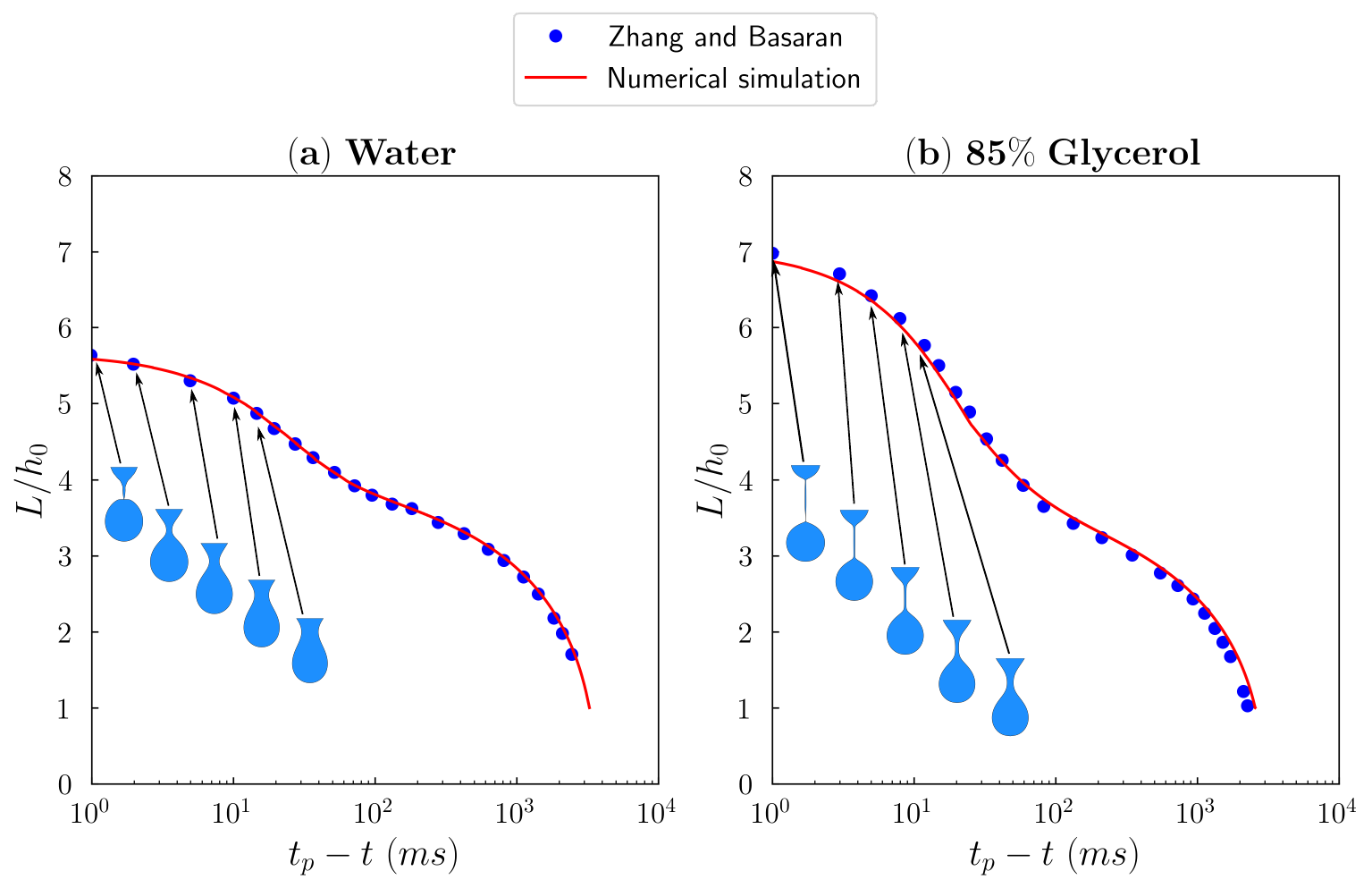}
     \caption{Evolution of non-dimensional length in the time approaching the pinch-off for (a) water and (b) $85\%$ glycerol. The inlet radius $h_0 = 0.0016$ m and inflow rate $=$ 1 mL/min.}
    \label{fig:L_validation}
\end{figure}

For the validation of our computational model, we use the experimental results by Zhang and Basaran~\cite{ZhangBasaran1995}. One crucial parameter to validate is the evolution of the droplet length with time. Because the calculation of the length is implicitly involved in the numerical model as explained in Algorithm~\ref{algo:1}. Figure~\ref{fig:L_validation} illustrates the comparison of the numerical simulation result with the experimental data. The length ($L$) is non-dimensionalized by the inlet radius ($h_0$). The time axis shows the time distance from the pinch-off. The comparison presented is for water and ($85\%$) glycerol solution. The viscosity, density, and surface tension are given in Table~\ref{table:1}. Initially, the droplet evolves slowly because the surface tension force is stronger than the effect of gravity. As we add more fluid, the drop becomes heavier and eventually, the gravitational force surpasses the surface tension. The surface tension starts to decrease the radius from the middle section, trying to minimize the surface energy. The length evolves much faster after the necking begins, suggesting an increase in the advection due to increasing velocity gradients. The primary droplet is separated when the radius is zero. The visualization of the droplet evolution after necking is also shown for water and ($85\%$) glycerol. Each droplet profile is attached to the point in the plot that corresponds to the time away from the pinch-off. The glycerol solution shows more elongation approaching the pinch-off time, which explains the effect of viscosity. The strong viscous effect allows glycerol droplet to have a long neck. In case of the water droplet, the surface tension force is much more dominant compared to the viscous forces approaching the singularity, resulting a shorter neck.
\begin{figure}[t]
    \centering
    \includegraphics[width=0.7\textwidth]{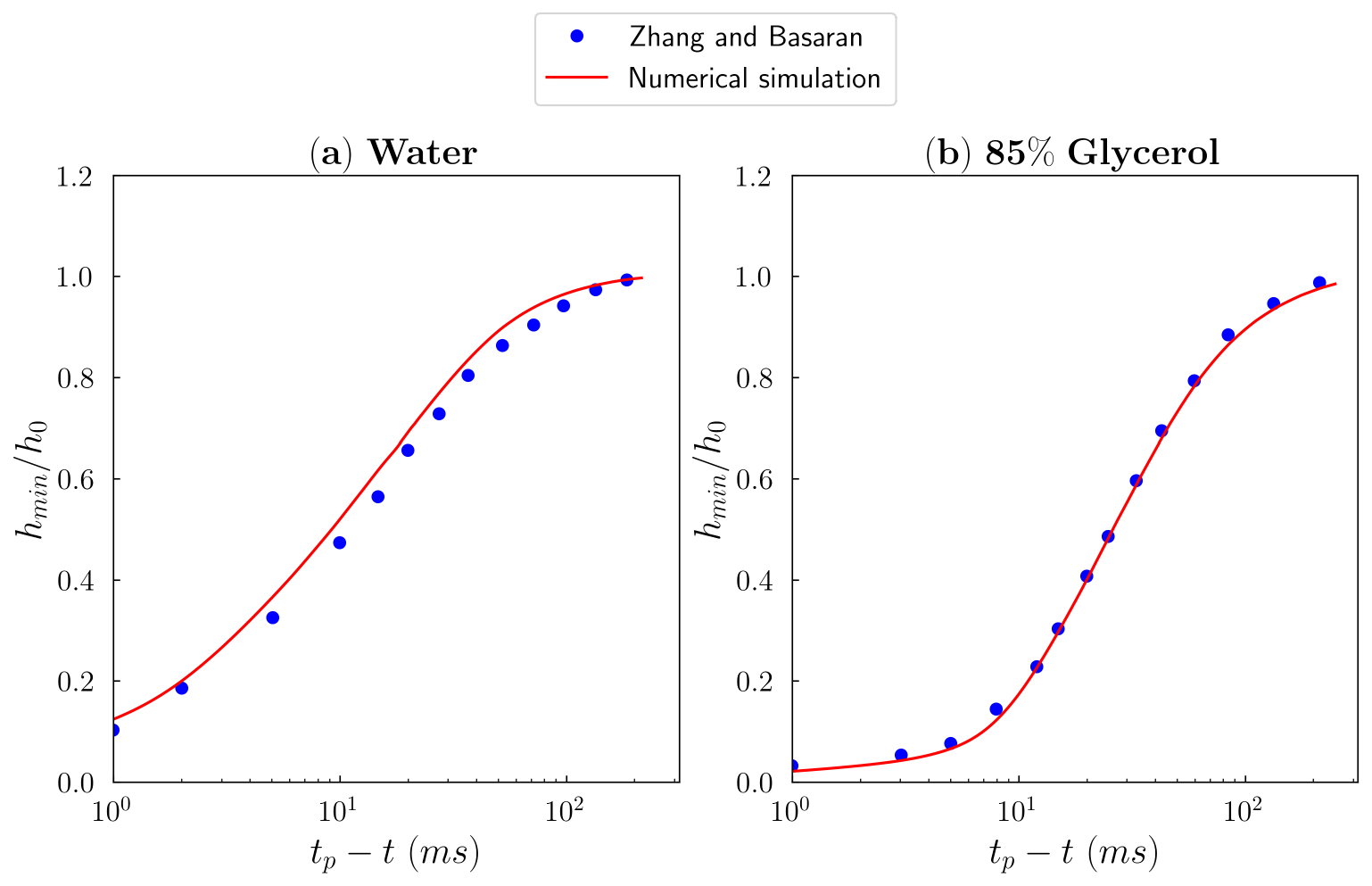}
     \caption{Evolution of non-dimensional radius of the narrowest location in time approaching the pinch-off for (a) water and (b) $85\%$ glycerol. The inlet radius $h_0 = 0.0016$ m and inflow rate $=$ 1 mL/min.}
    \label{fig:h_validation}
\end{figure}
Another important parameter for validation is the evolution of a minimum radius in time away from the pinch-off. Figure~\ref{fig:h_validation} illustrates numerical results for the evolution of this parameter, compared  with the experimental profiles for water and ($85\%$) glycerol solution. The profiles for both water and glycerol show similar evolution until a point where the dynamics start to become self-similar. From this point, the glycerol profile shows the influence of high viscosity by more elongation. The minimum radius decreases slower, which is evident by the long tail at the end of the profile. For the water droplet, the radial shrinkage is much faster due to the small viscosity compared to the glycerol solution. The numerical profile for glycerol agrees with the experiments very well. For the water, the profile shows a small amount of delay between the numerical and experimental results. This is due to the initially added artificial viscosity to increase diffusive behavior, which is inspired by the computational model by Eggers and Dupont~\cite{EggersDupont1994}. We increase the viscosity value initially for the low-viscosity materials since their highly convective nature introduces surface fluctuations. We reduce this added diffusion to zero well before the necking begins. Hence, the numerical profile for the water droplet starts to agree with the experiments right where the necking begins. This added diffusion shows no impact on the length evolution or the pinch-off location at all.
\subsection{Pinch-off dynamics of paraffin wax}
Paraffin wax is explored as one of the potential candidate fuels for hybrid propellant rockets~\cite{KarabeyogluAltmanCantwell2002}. In the combustion chamber of a hybrid rocket, a solid paraffin wax form a liquid layer on its burning surface. This liquid layer, under a high shear forcing, shows hydrodynamic instabilities that lead to the formation of droplets. These droplets are then entrained in the flow. Here, we explore the pinch-off dynamics of a pendant paraffin wax droplet using our computational model.
\begin{figure}[t]
    \centering
    \includegraphics[width=0.7\textwidth]{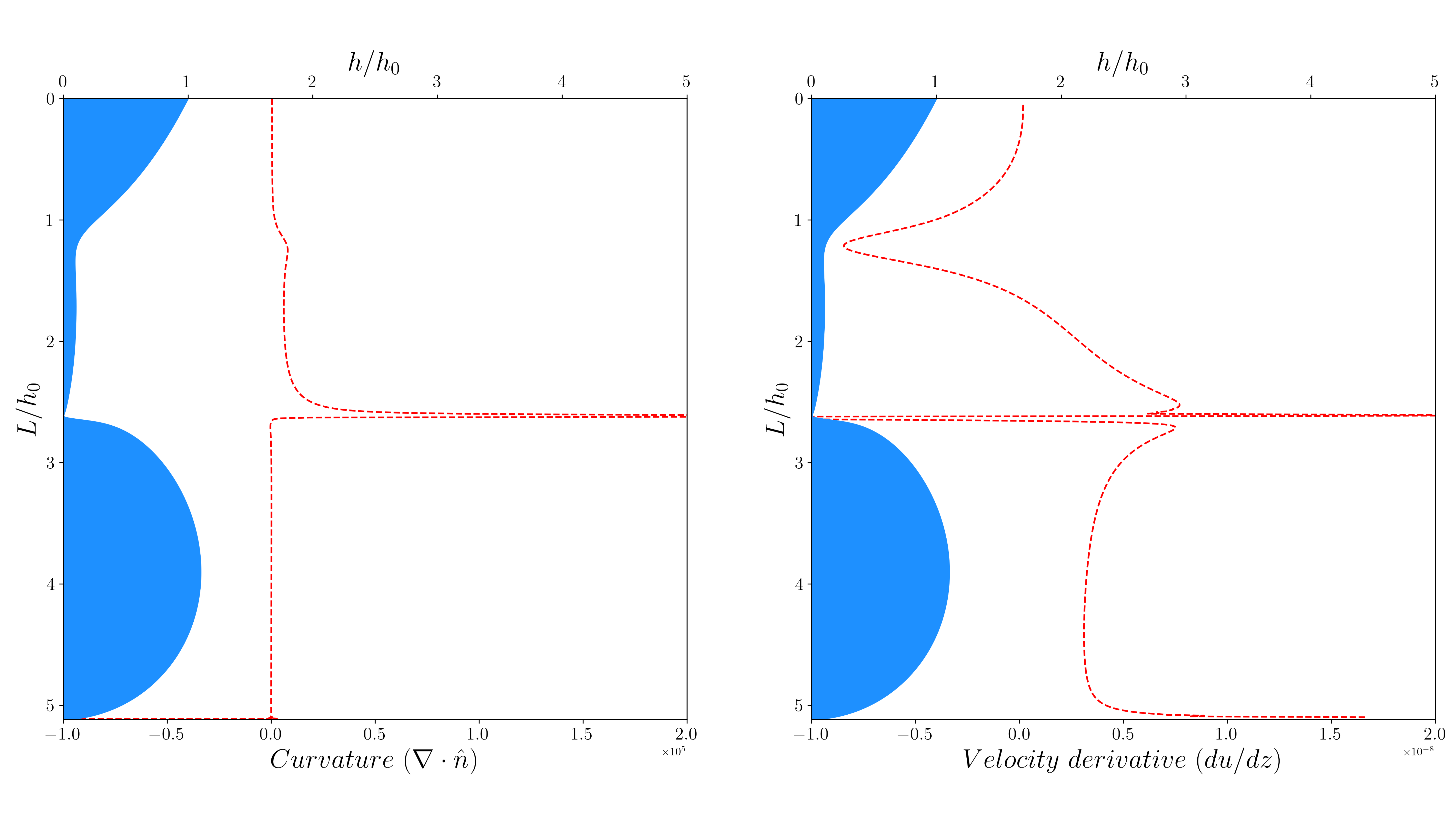}
    \caption{The curvature profile and the velocity derivative profile when the pinch-off happens. The inlet radius $h_0 = 0.0016$ m and inflow rate $=$ 1 mL/min. }
    \label{fig:velocity_curve_wax}
\end{figure}
The curvature is a chief attribute in understanding the pinch-off dynamics in droplet formation. Figure~\ref{fig:velocity_curve_wax} shows a paraffin wax droplet profile at pinch-off alongside a curvature profile. Approaching the singularity, the curvature starts to increase. But a finite time curvature blow-up happens at the pinch-off location, where the radius is zero. As explained in the introduction section, the fluid in close vicinity of this singular point is driven by very high-velocity gradients. The motion is independent of the initial and boundary conditions. This feature is mathematically described by a self-similar solution. The velocity derivative profile on the right in Fig.~\ref{fig:velocity_curve_wax} shows high gradients close to the singularity. Moreover, the velocity gradients change signs at the singular point location. This suggests that the fluid above and below the pinch-off point moves in opposite directions very quickly. This recoil provokes surface instabilities leading to the satellite drop formation.

However, it is also evident that the fluid motion approaching the pinch-off is highly convective in nature. The finite element model can approximate the solution but when the truncated terms are getting larger, the solution becomes unstable. Hence, the numerical scheme is augmented with a stabilization technique when the fluid viscosity is low. There are many options to consider for stabilization~\cite{DoneaHuerta2003}. We used the Streamline Upwinding (SU) and Streamline Upwinding Petrov Galerkin (SUPG) and found that this 1D problem can become stable with just the SU method. The SUPG method is better suited for problems with cross-convection in 2D or 3D. Also, the SUPG method regularizes the strong form residual that contains second-order derivatives, which can be problematic for $C^0$ continuous elements. However, the SU scheme just adds an artificial diffusion into the system and is over-diffusive in nature~\cite{BrooksHughes1982}. Therefore, we also decrease the artificial diffusivity as we refine the mesh adaptively to avoid adding too much diffusion. This stabilization was only enabled for low-viscosity fluids, like paraffin wax, water, etc. High-viscosity fluids like glycerol can be handled without any stabilization.

\section{Conclusion}
A  one-dimensional numerical model  is reliable for simulating droplet formation. The present model is validated using experimental pendant drops of water and glycerol, which were excellently matched with simulation results. The SU stabilization method was implemented since the low-viscosity fluids, like water and paraffin wax are highly convective. The velocity and curvature profile was shown for paraffin wax at the pinch-off time. The velocity was approaching infinity at the pinch-off location with the opposite signs suggesting that the pinch-off is followed by the recoil and then satellite drop formation.

The current computational model can be extended to capture the motion after pinch-off and satellite drops as well. Understanding the volume of satellite droplets is useful in accurately predicting the regression rate of the fuel in hybrid rockets. Also, the present model can be improved for droplet formation in a shear environment. The droplet formation in a turbulent environment may not be axisymmetric since the turbulence sets the initial droplet profile~\cite{RuthMostertPerrardDeike2019}. For non-symmetric interfaces, where no swirling motion assumption fails, two- or three-dimensional approaches are better, which is also recommended by Ambraneswaran~\cite{AmbravaneswaranWilkesBasaran2002}.  However, the droplet shape may not be assumed to be axisymmetric, the solution in the singularity region is still self-similar and follows the analogous dynamics. 
\section{Acknowledgement}
Funded by the United States Department of Energy’s (DoE) National Nuclear Security Administration (NNSA) under the Predictive Science Academic Alliance Program III (PSAAP III) at the University at Buffalo, under contract number DE-NA0003961.

This work was partially supported by the Department of Energy Office of Science Award DE-AC02-0000011838.
\printbibliography
\end{document}